\documentclass[12pt]{article}
\usepackage{amssymb,amsmath,cite,geometry,graphicx,url}
\usepackage[toc,page]{appendix}
\usepackage{mathtools}
\usepackage[tikz]{mdframed} 
\usetikzlibrary{decorations.pathmorphing}
\catcode`\@=11

\@addtoreset{equation}{section}
\def\theequation{\arabic{section}.\arabic{equation}}

\catcode`\@=11
\def\thesection{\arabic{section}.}

\def\appendix{\setcounter{section}{0}
        \def\thesection{Appendix.}
        \def\theequation{\Alph{section}.\arabic{equation}}}
\def\section{\@startsection{section}{1}{\z@}{3.5ex plus 1ex minus
   .2ex}{2.3ex plus .2ex}{\large\bf}}

\long\def\@makefntext#1{\parindent 0cm\noindent
\hbox to 1em{\hss$^{\@thefnmark}$}#1}

\newcommand{\captionfonts}{\small}
\makeatletter  
\long\def\@makecaption#1#2{%
  \vskip\abovecaptionskip
  \sbox\@tempboxa{{\captionfonts #1: #2}}%
  \ifdim \wd\@tempboxa >\hsize
    {\captionfonts #1: #2\par}
  \else
    \hbox to\hsize{\hfil\box\@tempboxa\hfil}%
  \fi
  \vskip\belowcaptionskip}
\makeatother   

\newcommand{\rd}{\mathrm{d}}
\newcommand{\rexp}{\mathrm{e}}
\newcommand{\ri}{\mathrm{i}}

\newcommand{\subrm}[1]{_{\mathrm{#1}}}
\DeclareMathSymbol{\Zeta}{\mathalpha}{operators}{"5A}

\begin{document}
\begin{titlepage}
\vspace{.5in}
\begin{flushright}
September 2017\\  
\end{flushright}
\vspace*{.5in}
\begin{center}
{\Large\bf
 Suppression of non-manifold-like sets\\[.5ex]
in the causal set path integral}\\  
\vspace{.4in}
{S.~P.~L{\sc oomis}\footnote{\it email: sloomis@ucdavis.edu} and
S.~C{\sc arlip}\footnote{\it email: carlip@physics.ucdavis.edu}\\
       {\small\it Department of Physics}\\
       {\small\it University of California}\\
       {\small\it Davis, CA 95616}\\{\small\it USA}}
\end{center}

\vspace{.5in}
\begin{center}
{\large\bf Abstract}
\end{center}
\begin{center}
\begin{minipage}{4.6in}
{\small
While it is possible to build causal sets that approximate spacetime manifolds, 
most causal sets are not at all manifold-like.  We show that a Lorentzian path 
integral with the Einstein-Hilbert action has a phase in which one large class of 
non-manifold-like causal sets is strongly suppressed, and suggest a direction
for generalization to other classes.  While we cannot yet show our argument 
holds for all non-manifold-like sets, our results make it plausible that the 
path integral might lead to emergent manifold-like behavior with no 
need for further conditions.
}
\end{minipage}
\end{center}
\end{titlepage}
\addtocounter{footnote}{-2}

\section{Introduction}

The causal set program offers a simple, elegant picture of spacetime as a discrete
set of points, characterized solely by their causal relations.  For all its elegance,
though, causal set theory has a potentially fatal flaw.  We know how to construct
causal sets that approximate spacetime manifolds, by starting with a manifold and
extracting a Poisson ``sprinkling'' of points.  But such manifold-like sets are highly
atypical; almost all causal sets do not look like any manifold at all.  If causal
sets are fundamental, and manifold-like behavior is emergent, a dynamical
process must somehow suppress almost all typical causal sets, leaving only the rare
manifold-like ones.  Finding such a process---especially one that has not been
artificially constructed merely to achieve this goal---is not easy.

In this paper, we show that the ordinary path integral with the causal set version of
the (Lorentzian) Einstein-Hilbert action has a phase in which one large class of
non-manifold-like causal sets is strongly suppressed.  The class for which we
can rigorously show this suppression, the two-level orders, is itself not ``typical''---we
certainly do not claim to show that \emph{all} non-manifold-like sets are suppressed.
But the two-level orders form a fairly large class, one much larger than the class of
manifold-like causal sets.  As we discuss in the conclusion, there are also
hints that our methods may extend to more general classes.  Our results thus make
it more plausible that the ordinary path integral, with no additional assumptions,
may be enough to lead to emergent manifold-like behavior.

A numerical analysis of two-dimensional causal sets has shown a similar transition
between a phase dominated by non-manifold-like causal sets and one dominated
by manifold-like sets \cite{Glaser16}.  In one way, that result is stronger than ours,
since it accounts for all non-manifold-like sets.  On the other hand, our results
are analytic, hold in any dimension, and use the Lorentzian path integral rather
than analytically continuing to Riemannian signature.

\section{Non-manifold-like causal sets}

The procedure for constructing a manifold-like causal set is well understood
\cite{Bombelli87}.  One starts with a finite-volume region of a manifold with a
Lorentzian metric, ``sprinkles'' points randomly by a Poisson process,
determines the causal relations among these points from the causal structure
of the manifold, and then ``forgets'' the manifold, keeping only the points
and their causal relations.  For a dense enough sprinkling of points, the
resulting causal set retains the fundamental properties of the original manifold:
the Alexandrov neighborhoods determine the topology, the causal relations
determine the conformal class of the metric, and the density of points
determines the conformal factor \cite{Bombelli89,Major}.

But such manifold-like causal sets are highly atypical.  The ``typical''
causal set is a Kleitman-Rothschild (KR) order, a three-level causal set with
approximately $n/4$ points in the ``bottom'' and ``top'' layers and $n/2$
points in the ``middle'' layer \cite{Kleitman75}.  In fact, as $n\rightarrow\infty$,
the proportion of $n$-element causal sets that are KR orders goes to one.

Many other non-manifold-like causal sets also occur frequently.  There is,
in fact, a hierarchy of classes of non-manifold-like causal sets
\cite{Dhar, Kleitman79, Promel, Henson}.  Each class is characterized by a
parameter $p$ that is the proportion of possible relations that are
actualized, and is dominated by causal sets with a particular number of
levels.  The dependence of the size of the class on $p$ is not smooth,
but is described by an piecewise continuous function with infinitely many
``phase transitions'' characterized by either the creation of new layers or
changes in the relative sizes of the layers. The intricacies of these classes
are beyond the scope of this paper---see \cite{Promel} for details---but it is
sufficient to point out that the class of non-manifold-like causal sets is
dominated by three-level orders, primarily the KR orders, followed by
two-level orders and then four-level orders.

In this paper we will focus on the simplest case of two-level orders.
Though these are not as dominant as the three-level orders, they still form
a significant part of the collection of non-manifold like causal sets.

\section{Causal set path integrals}

To define a path integral for causal sets, we need two ingredients: an appropriate
generalization of the Einstein-Hilbert action and a discrete version of an integration
measure.  The action we shall use, the Benincasa-Dowker action, was introduced
in \cite{Benincasa}.  For a causal set $C$ with $n$ elements, it takes the general form
\cite{Dowker,Glaser14}
\begin{equation}\label{action}
  \frac{1}{\hbar}S(C) = \mu\left(n + \sum_{k=0}^{k_{\mathrm{max}}}\lambda_k N_k\right)
\end{equation}
where $\mu$ and $\lambda_k$ are appropriately chosen parameters and
$N_k$ denotes the number of pairs of elements $\{x,y\}\subset C$ such that the
cardinality of the set $\{z\in C:x\prec z \prec y\}$ is equal to $k$.  The upper limit
$k_{\mathrm{max}}$ can be finite or infinite, though it has a lower bound of
$\lfloor{2+\frac{d}{2}}\rfloor$, where $d$ is the target spacetime dimension.

Eq.~\eqref{action} replicates the Einstein-Hilbert action in the following sense.  Suppose
we construct a causal set by Poisson sprinkling points into a manifold of the target dimension.
Then for a high enough sprinkling density and the correct choices of $\mu$ and $\lambda_k$,
$S(C)$ is equal to the Einstein-Hilbert
action on average. The specific definitions of $\mu$ and $\lambda_k$ are complicated,
but for $d=4$ and $k\subrm{max}=3$ we have
\begin{equation}\label{4d}
\frac{1}{\hbar}\mathcal{S}(C) = \left(\frac{l}{l_p}\right)^2\left(n - N_0 + 9N_1 -16N_2 +8N_3\right)
\end{equation}
where $l_p$ is the Planck length and $l$ is a length scale determined by the sprinkling
density of events into the spacetime.

For our ``integration measure'' we shall simply sum over causal sets.  As in causal dynamical
triangulations \cite{Loll}, we should perhaps include a combinatorial weight to avoid
overcounting causal sets with special symmetries, but that will not affect our conclusions.
The Lorentzian partition function over any particular class $\mathcal{C}$ of causal sets is then
\begin{equation}\label{partition}
\mathcal{Z}[\mu,\lambda_0] = \sum_{C\in\mathcal{C}}\exp\left(\frac{\ri}{\hbar}S(C)\right)
  = \sum_{C\in\mathcal{C}}\exp\left({\ri\mu}%
   \left[n + \sum_{k=0}^{k_{\mathrm{max}}}\lambda_k N_k\right]\right)
\end{equation}
We will be interested in the large $n$ behavior of this quantity; for a manifold-like causal set 
with a fixed sprinkling density, this is the large volume limit.

\section{Suppression of two-level orders \label{secx}}

For this paper we focus on two-level orders, that is, causal sets $C$ of size $n$ such that
there are no three distinct elements $x,y,z\in C$ satisfying $x\prec y\prec z$.  This means
that $N_k = 0$ for $k>0$.  Intuitively, such sets have only two ``moments of time,'' and
clearly do not resemble manifolds. As we have mentioned, while they are less common
than the three-level KR orders, two-level orders are still much more common than
manifold-like causal sets, and they threaten to dominate the path integral.

For any $n$-element causal set, $N_0$ can be no larger than $N_{\mathrm{max}}%
=\frac{n(n-1)}{2}$.  We classify such sets by the proportion $0\leq p\leq 1$ of relations,
given by $N_0 = pN_{\mathrm{max}}$.  For fixed $n$, $p$ is a discrete parameter, but in
the limit of large $n$ we can approximate it as continuous.   The utility of this classification
is that the Benincasa-Dowker action is constant over the class of two-level sets with a
fixed $p$.  Denoting such a class by $\mathcal{C}_{p,n}$, we can write the partition
 function over two-level orders of size $n$ as
\begin{equation}
  \mathcal{Z}[\mu,\lambda_0] =\int\rd p \,|\mathcal{C}_{p,n}| \rexp^{\ri S(p)/\hbar}
  = \rexp^{\ri\mu n} \int_{0}^1\rd p\, \left|\mathcal{C}_{p,n}\right|%
  \exp\left(\frac{1}{2}\ri\mu\lambda_0 pn^2+\mathrm{o}(n^2)\right)
\label{x2}
\end{equation}
where $|\mathcal{C}_{p,n}|$ is the cardinality of the class $\mathcal{C}_{p,n}$.  Here
we have written $N_\mathrm{max} = \frac{1}{2}n^2+\mathrm{o}(n^2)$, where
$\mathrm{o}(n^2)$ denotes terms subleading to $n^2$, which will be negligible in
the large $n$ limit.

To calculate $|\mathcal{C}_{p,n}|$ we consider a decomposition into classes
$\mathcal{C}_{q,p,n}$ where we put $qn$ of the elements in the ``top'' level and
$(1-q)n$ in the ``bottom'' level.  Let us denote $\alpha = q(1-q)$, where
$\alpha \leq \frac{1}{4}$ since $0\leq q\leq 1$.  From the structure of the system,
there can be at most $\alpha n^2$ relations---the maximum occurs when every
``bottom'' element is related to every ``top'' element---so from the definition of
$p$, we have $\alpha \geq \frac{1}{2}p$.  This in turn implies that $p \leq \frac{1}{2}$.

The number of ways to choose $p N_{\mathrm{max}}= \frac{1}{2}pn(n-1)$ pairs
from the possible $\alpha n^2$ relations is
\begin{equation}
  |\mathcal{C}_{q,p,n}| = \binom{\alpha n^2}{\frac{1}{2}p n(n-1)}
\end{equation}
With both arguments large, we can expand the binomial as
\begin{align}
  \begin{split}
    \ln |\mathcal{C}_{q,p,n}|
    =& \alpha n^2\ln(\alpha n^2)-\frac{1}{2}p n^2\ln\left(\frac{1}{2}pn^2\right)
    - \left(\alpha -\frac{1}{2}p\right)n^2\ln\left(\left[\alpha-\frac{1}{2}p\right]n^2\right)
    +\mathrm{o}(n^2)\\
    =& \left[\alpha\ln\alpha - \frac{1}{2}p\ln\left(\frac{1}{2}p\right)
    - \left(\alpha-\frac{1}{2}p\right) \ln\left(\alpha-\frac{1}{2}p\right) \right]n^2
    +\mathrm{o}(n^2)
  \end{split}
\end{align}
For $\frac{1}{2}p\leq \alpha \leq \frac{1}{4}$, this is is a monotonically increasing function
of $\alpha$. This means that $|C_{q,p,n}|$ is maximized for $q = \frac{1}{2}$.  Now,
$\left|\mathcal{C}_{p,n}\right| $ is bounded by
\begin{equation}
  \left|\mathcal{C}_{\frac{1}{2},p,n}\right| \leq |\mathcal{C}_{p,n}| \leq \sum_q |\mathcal{C}_{q,p,n}|
\end{equation}
In the large $n$ limit, the upper bound is dominated by the maximal value of $q$, so
\begin{equation}
  \ln|\mathcal{C}_{p,n}| =\ln|\mathcal{C}_{\frac{1}{2},p,n}|+\mathrm{o}(n^2)
  =\frac{1}{4}h(2p)n^2 +\mathrm{o}(n^2)\quad \left(p\leq \frac{1}{2}\right)
\label{x1}
\end{equation}
where 
\begin{equation}
h(x) = -x\ln x - (1-x)\ln(1-x)
\label{xyz}
\end{equation} 
is the entropy function.
(As we saw above, $p\le\frac{1}{2}$ for two-level sets, so $|\mathcal{C}_{p,n}| =0$
for $p > \frac{1}{2}$.)

Using (\ref{x1}), we can write the partition function as
\begin{equation}
  \mathcal{Z}[\mu,\lambda_0] = \rexp^{\ri\mu n} \int_{0}^{1/2}\rd p \,
  \exp\left(\frac{1}{2}\mathrm{i}\mu\lambda_0 pn^2+\frac{1}{4}h(2p)n^2
  +\mathrm{o}(n^2)\right)
\label{x3}
\end{equation}
To simplify notation, we define 
\begin{equation}
-\frac{\mu\lambda_0}{2} = \beta, \quad 
2p = x 
\label{xxa1}
\end{equation}
Note that $0\le x\le 1$ and that, from (\ref{4d}), $\beta>0$.  The exponent 
in (\ref{x3}) is then $\frac{n^2}{4} E(x)$, with
\begin{equation}
E(x) = -2i\beta x + h(x)
\label{xxa2}
\end{equation}

We will evaluate the integral by the method of steepest descents.\footnote{An
earlier attempt to determine the integral in a quadratic approximation failed;
we thank Lisa Glaser for pointing out an algebraic error that invalidated our
first approach.}   Here we sketch the method and results; details are given 
in the appendix.   We first find the saddle point:
\begin{align}
E'(x) &= 0 = -2i\beta -\ln x + \ln(1-x) \label{xxa3} \\
&\Rightarrow\  x_0 = \frac{e^{-i\beta}}{2\cos\beta} = \frac{1}{2}(1-i\tan\beta), \ \ 
   1-x_0 =  \frac{e^{i\beta}}{2\cos\beta} = \frac{1}{2}(1+i\tan\beta) \nonumber
\end{align}
The second derivative at $x=x_0$ is
\begin{equation}
E^{\prime\prime}(x_0) = -\frac{1}{x_0} - \frac{1}{1-x_0} = -4\cos^2\beta 
\label{xxa4}
\end{equation}
so the direction of steepest descent is $x-x_0$ real.  At the saddle point,
\begin{align}
h(x_0)   &= -\frac{e^{-i\beta}}{2\cos\beta}\ln\left(\frac{e^{-i\beta}}{2\cos\beta}\right) 
              -\frac{e^{i\beta}}{2\cos\beta}\ln\left(\frac{e^{i\beta}}{2\cos\beta}\right)   
            = \beta\tan\beta + \ln(2\cos\beta)  \nonumber\\[.5ex]
E(x_0) &= -2i\beta x_0 + h(x_0) = -i\beta + \ln(2\cos\beta) 
 \label{xxa5}
\end{align} 
Remembering that the exponent is $\frac{n^2}{4} E(x)$, we have a saddle point contribution
of
\begin{equation}
{\mathcal Z}[\mu,\lambda_0] \sim \frac{e^{i\mu n}}{n}
    {\sqrt{\frac{\pi}{2|E^{\prime\prime}(x_0)|}}}\exp\left\{\frac{n^2}{4} E(x_0)\right\}
    = \sqrt{\frac{\pi}{8}}\,\frac{e^{i\mu n}}{n\cos\beta}
   \exp\left\{\frac{n^2}{4}\left[-i\beta + \ln(2\cos\beta)\right]\right\} 
\label{xxa6}
\end{equation}
If $|\cos\beta|<\frac{1}{2}$, the real part of the exponent is negative, and the
path integral is exponentially suppressed.

\begin{figure}
\begin{center}
\begin{tikzpicture}
\draw[decorate,decoration={zigzag, segment length=2mm, amplitude=.4mm},very thick]%
     (4,0) -- (7,0); 
\node at (7,0) {$\bullet$};
\draw[decorate,decoration={zigzag, segment length=2mm, amplitude=.4mm},very thick]%
     (11,0) -- (14,0); 
\node at (11,0) {$\bullet$};
\draw[dashed] (7,0) -- (7,-2.25);
\draw[dashed] (7,-2.25) -- (11,-2.25);
\draw[dashed] (11,0) -- (11,-2.5);
\node at (9,-2.25) {$\bullet$};
\node at (9,-2.55) {$\scriptstyle x_-$};
\node at (7.2,0) {$\scriptstyle 0$};
\node at (10.8,0) {$\scriptstyle 1$};
\node at (6.7,-1.3) {$\scriptstyle C_1^-$};
\node at (11.35,-1.3) {$\scriptstyle C_2^-$};
\node at (8.25,-2) {$\scriptstyle C_3^-$};
\end{tikzpicture}
\end{center}
\caption{Deformed contour through the saddle point at $x_0$ 
with $\tan\beta>0$ \label{fig1}}
\end{figure}
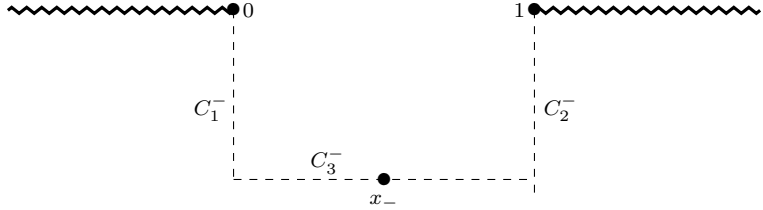

This is not quite the whole story.  The method of steepest descent requires   
a contour deformation, and we must check that the rest of the contour  
does not spoil the result.  For $\tan\beta>0$, the
saddle point is in the lower half plane, and the contour is shown in figure
\ref{fig1}.  We show in the appendix that the remaining pieces of the
contour, $C_1^-$ and $C_2^-$, are also exponentially suppressed.  If,
on the other hand, $\tan\beta<0$, we must deform the contour into the
 upper half plane, and the remaining pieces are \emph{not} suppressed.
We thus conclude that the path integral for two-level orders is exponentially
suppressed at large volume provided that
\begin{equation}
\tan\left(-\frac{\mu\lambda_0}{2}\right)>0\quad \text{and}\quad
\left|\cos\frac{\mu\lambda_0}{2}\right|<\frac{1}{2} \quad
\Rightarrow\ \tan\left(-\frac{\mu\lambda_0}{2}\right)>\sqrt{3}
\label{xxa7}
\end{equation}

We can also carry the analysis one step further.  The saddle point
approximation (\ref{xxa6}) is not exact, and one might worry about the
higher order terms in the exponent.  In the appendix, we give a
rigorous bound: exponential suppression is guaranteed provided that
\begin{equation}
\tan\left(-\frac{\mu\lambda_0}{2}\right) 
    >\left(\frac{27}{4}e^{-1/2} - 1\right)^{1/2} \approx 1.759
\label{xxa8}
\end{equation}
This gives a minimum value of $|\mu\lambda_0| \approx 2.108$, or
a scale $\ell \approx 1.452\ell_p$ in (\ref{4d}).
 
\section{Discussion}
The program we have described can be summarized as follows:
\begin{enumerate}
\item Identify a class of causal sets that can be divided into subclasses characterized by some
parameters $p_i$ such that the action is constant over each subclass.
\item Count how large each subclass is, to leading order in the size $n$ of the set, as a
function of the parameters $p_i$.
\item Analytically evaluate the partition function as an integral over $p_i$, and study how it
depends on the parameters $\mu$ and $\lambda_i$ in the action.
\end{enumerate}

We have carried this out for a particularly simple case, in which the division into easily countable
subclasses was fairly straightforward.  But there are hints that our results can be generalized.
Once we move beyond two-level orders, the action (\ref{4d}) will include contributions from
$N_1$, $N_2$, and $N_3$, greatly complicating the counting.  But for sets with only a few
levels, these contributions may be strongly suppressed.

Consider, for example, a KR order, which has approximately $n/4$ points in a ``bottom'' level,
$n/2$ in a ``middle'' level, and $n/4$ in a ``top'' level.\footnote{More precisely \cite{Kleitman75},
a KR order has between $n/4 - n^{1/2}\ln n$ and $n/4 + n^{1/2}\ln n$ points in the bottom
and top levels, and between $n/2 - \ln n$ and $n/2 + \ln n$ points in the middle level.}
Pick a ``bottom'' point $x$ and a ``top'' point $y$.  Typically, $x$ will link to approximately
$n/4$ points in the middle level.  Imagine coloring these points red, and the remaining points
blue.  For $\{x,y\}$ to contribute to $N_1$, $y$ must then link to exactly one red point in the
middle level, along with approximately  $n/4$ blue points.  It is easy to see that the probability
of such a pattern goes as a polynomial in $n$ times $2^{-n/2}$.  The same is true for
contributions to $N_2$ (for which $y$ must link to exactly two red points) and $N_3$ (for
which $y$ must link to exactly three red points).  Similar arguments should hold whenever
the number of levels is small.

This suppression should reduce the analysis of KR orders, and perhaps similar few-level sets,
to the form we have already considered, in which only $N_0$ is important.  This is still a 
preliminary argument, of course.  The $N_0$ combinatorics will be different for
different orders, and one must check that the ``atypical'' few-level causal sets---three-level
sets with different distributions of points or relations from the KR orders, for instance---remain
subdominant.  Here the combinatoric results of \cite{Promel} may prove useful, but much
more work is needed.   There are also subtleties involving the difference between labeled
and unlabeled causal sets that require careful attention \cite{Henson}.  Numerical exploration
of distributions of causal sets and relations may shed additional light on these problems.

\vspace{1.5ex}
\begin{flushleft}
\large\bf Acknowledgments
\end{flushleft}
We are very grateful to Lisa Glaser for pointing out a crucial error in an early
version of this work.  We also thank David Rideout for helpful conversations.
This work was supported in part by U.S.\ Department of Energy grant
DE-FG02-91ER40674.

\appendix
\section{Steepest descent details}

In this appendix we describe some of the details involved in the steepest descent 
calculation of section 4.\\[-.5ex]

\noindent{\bf Contours}\\[-.5ex]

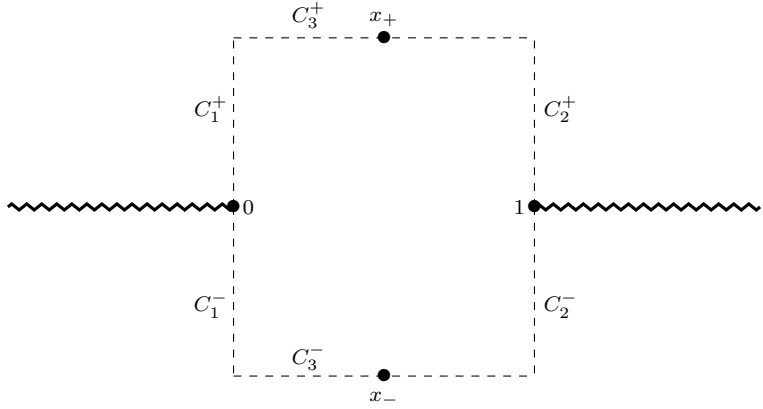
\begin{figure}
\begin{center}
\begin{tikzpicture}
\draw[decorate,decoration={zigzag, segment length=2mm, amplitude=.4mm},very thick]%
     (4,0) -- (7,0); 
\node at (7,0) {$\bullet$};
\draw[decorate,decoration={zigzag, segment length=2mm, amplitude=.4mm},very thick]%
     (11,0) -- (14,0); 
\node at (11,0) {$\bullet$};
\draw[dashed] (7,0) -- (7,-2.25);
\draw[dashed] (7,-2.25) -- (11,-2.25);
\draw[dashed] (11,0) -- (11,-2.25);
\node at (9,-2.25) {$\bullet$};
\node at (9,-2.55) {$\scriptstyle x_-$};
\node at (7.2,0) {$\scriptstyle 0$};
\node at (10.8,0) {$\scriptstyle 1$};
\node at (6.7,-1.3) {$\scriptstyle C_1^-$};
\node at (11.35,-1.3) {$\scriptstyle C_2^-$};
\node at (8,-2) {$\scriptstyle C_3^-$};
\node at (9,2.25) {$\bullet$};
\node at (9,2.5) {$\scriptstyle x_+$};
\draw[dashed] (7,0) -- (7,2.25);
\draw[dashed] (11,0) -- (11,2.25);
\draw[dashed] (7,2.25) -- (11,2.25);
\node at (6.7,1.3) {$\scriptstyle C_1^+$};
\node at (11.35,1.3) {$\scriptstyle C_2^+$};
\node at (8,2.55) {$\scriptstyle C_3^+$};
\end{tikzpicture}
\end{center}
\caption{Deformed contours through the saddle point at $x_0$ \label{fig2}}
\end{figure}

The integral (\ref{x3}) is over the interval $0<x<1$.  For the method of steepest 
descent, we must first deform the contour to go through the saddle point in
the direction of steepest descent.  The saddle point is $x_0 = \frac{1}{2}(1-i\tan\beta)$
and the direction of steepest descent is $x-x_0$ real, so the contours are those
of figure \ref{fig2}, where the lower branch is applicable for $\tan\beta>0$
and the upper for $\tan\beta<0$.  

Let us first exclude the contour in the upper half plane.  Consider $C_1^+$.  
We can write
\begin{equation}
x=+iw, \quad  0<w<\frac{1}{2}|\tan\beta|
\label{xxd1}
\end{equation}
from which, with the branch cuts shown in figure \ref{fig2},
\begin{equation}
\ln x = \frac{\pi i}{2} + \ln w, \quad \ln(1-x) = \ln\sqrt{1+w^2} - i\tan^{-1}w
\label{xxd2}
\end{equation}
with the inverse tangent lying between $0$ and $\frac{\pi}{2}$.   Hence
\begin{align}
h(x)  &= -x\ln x - (1-x)\ln(1-x)  \nonumber\\
        &= -iw\left(\frac{\pi i}{2} + \ln w\right) 
          - (1-iw)\left(\ln\sqrt{1+w^2} - i\tan^{-1}w\right) \nonumber\\
        &= \frac{\pi}{2}w + w\tan^{-1}w - \ln\sqrt{1+w^2} + \text{\it imaginary part}
\label{xxd3}
\end{align}
For $\Im x>0$, the contribution from the term $-2i\beta x$ in $E(x)$ is positive, and
\begin{equation}
\Re\,E = \left(\frac{\pi}{2} + \tan^{-1}w + 2\beta\right)w - \ln\sqrt{1+w^2}
\label{xxd4}
\end{equation}
For positive real $w$, this is always positive, so the integral acquires an 
exponentially large contribution from $C_1^+$.  This rules out 
the saddle point approximation for this contour.

Next  consider the contour in the lower half plane. On $C_1^-$, we can 
write
\begin{align}
&x = -iw, \quad 0\le w \le \frac{1}{2}\tan\beta   \nonumber\\
&\ln(-iw) = -\frac{\pi i}{2} + \ln w, \quad
\ln(1+iw) = \ln\sqrt{1+w^2} + i\tan^{-1}w
\label{xxb3}
\end{align}
with the inverse tangent again lying between $0$ and $\frac{\pi}{2}$.  Then
\begin{align}
h(x)  
        &= iw\left(-\frac{\pi i}{2} + \ln w\right) - (1+iw)\left(\ln\sqrt{1+w^2} + i\tan^{-1}w\right) \nonumber\\
        &= \frac{\pi}{2}w + w\tan^{-1}w - \ln\sqrt{1+w^2} + \text{\it imaginary part}
\label{xxb4}
\end{align}
and thus
\begin{equation}
\Re\,E = \left(\frac{\pi}{2} + \tan^{-1}w - 2\beta\right)w - \ln\sqrt{1+w^2}
\label{xxb5}
\end{equation}

For $\beta>\frac{\pi}{2}$, this is always negative, and the contribution from $C_1^-$ is
exponentially suppressed.  For $0<\beta<\frac{\pi}{2}$, the requirement that
$|\cos\beta|<\frac{1}{2}$ limits us to the range $\frac{\pi}{3}<\beta<\frac{\pi}{2}$.  To 
proceed, let us determine the maximum value of $\Re\,E$ in this range.

Note first that at $w=0$, $\Re\,E=0$ and the derivative
\begin{equation}
\frac{d(\Re\,E)}{dw} = \frac{\pi}{2} + \tan^{-1}w - 2\beta 
\label{xxb6}
\end{equation}
is negative, so $\Re\,E<0$ for small $w$.  The turning point occurs at
\begin{equation}
\frac{\pi}{2} + \tan^{-1}w - 2\beta  = 0 \ \Rightarrow \ w=-\cot2\beta 
    = \frac{1}{2}(\tan\beta - \cot\beta)
\label{xxb7}
\end{equation}
For $\frac{1}{2}(\tan\beta - \cot\beta) < w < \frac{1}{2}\tan\beta$, $\Re\,E$ is increasing,
so its maximum in this range will occur at the endpoint $w=\frac{1}{2}\tan\beta$.  At that
maximum,
\begin{equation}
\Re\,E = \frac{1}{2}\left[\frac{\pi}{2} + \tan^{-1}\left(\frac{1}{2}\tan\beta\right) - 2\beta\right]\tan\beta 
    - \ln\sqrt{1+\frac{1}{4}\tan^2\beta}
\label{xxb8}
\end{equation}
Treating this quantity as a function of $\beta$ and using Mathematica \cite{Math} 
to determine its zeros, we find that it is negative for $.9474<\beta<\frac{\pi}{2}$, 
an interval that includes the full range of interest.  Hence $\Re\,E(w)<0$ for any 
$\beta$ in the range $\frac{\pi}{3}<\beta<\frac{\pi}{2}$, and the contribution of
the contour $C_1^-$ is again exponentially suppressed.

The contour $C_2^-$ is basically a reflection, and gives the identical suppression.   Let
\begin{equation}
x = 1-iv , \quad 0\le v \le \frac{1}{2}\tan\beta
\label{xxc1}
\end{equation}
Then
\begin{align}
h(x)  
        &= -(1-iv)\left(\ln\sqrt{1+v^2} - i\tan^{-1}v\right) -iv \left(\frac{\pi i}{2} + \ln v\right)\nonumber\\
        &= \frac{\pi}{2}v + v\tan^{-1}v - \ln\sqrt{1+v^2} + \text{\it imaginary part}
\label{xxc2}
\end{align}
and
\begin{equation}
\Re\,E = \left(\frac{\pi}{2} + \tan^{-1}v - 2\beta\right)v - \ln\sqrt{1+v^2}
\label{xxc3}
\end{equation}
which exactly matches (\ref{xxb5}).  This match is not accidental; it follows from 
the fact that 
$$\Re h(1-iv) = \Re h(iv) = \Re h(-iv)$$
as long as we stay on the same branch of the logarithm.\\[-.5ex]

\noindent{\bf Error estimates}\\[-.5ex]

The integral (\ref{xxa6}) is based on a quadratic approximation to $E(x)$.  In 
this case, we can also get control over the errors.  Let $x = \frac{1}{2}(1- u)$.  
It is then easy to check that for $n\ge2$,
\begin{equation}
\frac{d^n h}{du^n} = -\frac{1}{2}\frac{(n-2)!\ }{(1-u)^{n-1}} -\frac{(-1)^n}{2}\frac{(n-2)!\ }{(1+u)^{n-1}} 
\label{z1}
\end{equation}
Now expand $E(x)$ around $x_0$.  Since $u_0 =  i\tan\beta$
is imaginary, the two terms in (\ref{z1}) evaluated at $x_0$ are complex 
conjugates; the odd derivatives are imaginary, while the even derivatives are real.  
The Taylor expansion for $E(x)$ around $x_0$, with $x-x_0$ real, is then
\begin{equation}
\Re E(x) =\Re  E(x_0) 
     -\sum_{n=1}^{\infty} \frac{2^{2n}}{2n(2n-1)}[\cos^{2n-1}\!\beta][\cos(2n-1)\beta]\,(x-x_0)^{2n}
\label{z2}
\end{equation}
where (\ref{xxa3}) has been used to evaluate $\Re (1-u_0)^{-(2n-1)}$.  Hence
\begin{align}
\bigl| \Re(E(x) - E(x_0) &+ 2\cos^2\beta(x-x_0)^2)\bigr| \nonumber\\
     &\le \sum_{n=2}^{\infty} \frac{2^{2n}}{2n(2n-1)}|\cos^{2n-1}\!\beta|
     |\cos(2n-1)\beta|\,(x-x_0)^{2n}
     \nonumber\\
     &\le \sum_{n=2}^{\infty} \frac{2^{2n}}{2n(2n-1)}\left(\frac{1}{2}\right)^{4n-1} = \
     \sum_{n=2}^{\infty} \frac{2^{-2n}}{n(2n-1)}
\label{z3}
\end{align}
using the facts that $|\cos\beta|\le\frac{1}{2}$ and $|x-x_0|\le\frac{1}{2}$.  
The sum evaluates
to 
$$\frac{3}{2}\ln\frac{3}{2} + \frac{1}{2}\ln\frac{1}{2} - \frac{1}{4} \approx 0.0116$$
We can thus state, for instance, that on the line $0<\Re x<1$, 
$\Im x = -\frac{i}{2}\tan\beta$ with $\tan\beta>0$---that is, the line through 
the saddle point $x_0$---the exponent $E(x)$ is negative as long as
\begin{equation}
|\cos\beta| <2\cdot3^{-3/2}e^{-1/4} \approx 0.4942
\label{z4}
\end{equation}
which in turn yields (\ref{xxa8}).  We do not know whether this is a sharp limit.

\end{document}